\documentclass{elsart}
\usepackage{epsfig}

\begin{document}

\begin{frontmatter}

\title{Rigidity-dependent cosmic ray energy spectra in the
knee region obtained with the GAMMA experiment}
\author[ARM]{A.P. Garyaka}, 
\author[ARM]{R.M. Martirosov},
\author[ARM]{S.V. Ter-Antonyan\corauthref{cor}},
  \ead{samvel@yerphi.am}	
\corauth[cor]{Corresponding author:}
\author[RUS]{N. Nikolskaya},
\author[MON]{Y.A. Gallant},
\author[USA]{L. Jones}
\author[FRA]{and J. Procureur}
\address[ARM]{Yerevan Physics Institute, 2 Alikhanyan Br. Str., 375036 
Yerevan, Armenia}
\address[MON]{
Laboratoire de Physique Th\'eorique et Astroparticules, Universit\'e
Montpellier II, CNRS/IN2P3, France}
\address[USA]{University of Michigan, Department of Physics, USA}
\address[RUS]{Moscow Lebedev Physics Institute, Russia}
\address[FRA]{Centre d'Etudes Nucl\'eaires de Bordeaux-Gradignan, 
Gradignan, France}
\begin{abstract}
On the basis of the extensive air shower (EAS) data obtained by the
GAMMA experiment, the energy spectra and elemental composition of the
primary cosmic rays are derived in the $10^3-10^5$ TeV energy
range. The reconstruction of the primary energy spectra is carried out
using an EAS inverse approach in the framework of the SIBYLL2.1 and QGSJET01 
interaction models and the hypothesis
of power-law primary energy spectra with 
rigidity-dependent knees.
The energy spectra 
of primary $H, He, O$-like and $Fe$-like nuclei obtained with the
SIBYLL interaction model agree with 
corresponding extrapolations of the balloon and satellite data  
to $\sim10^3$ TeV energies. 
The energy spectra obtained from the QGSJET model show a predominantly 
proton composition in the knee region. The 
rigidity-dependent knee feature of the primary energy spectra for each
interaction model is displayed at the following rigidities: 
$E_R\simeq2500\pm200$ TV (SIBYLL) and $E_R\simeq3100-4200$ TV 
(QGSJET).\\
All the results presented are derived taking into account the detector 
response, the
reconstruction uncertainties of the EAS parameters, and fluctuations
in the EAS development.  
\end{abstract}
\end{frontmatter}

\section{Introduction} 
The investigation of the energy spectra and elemental composition
of primary cosmic rays in the knee region ($10^3-10^5$ TeV)
remains one of the intriguing  problems of 
modern high energy cosmic-ray physics. 
Despite the fact that these investigations
have been carried out for more 
than half a century, the data on the
elemental primary energy spectra at energies $E>10^3$ TeV 
still need improvement.
The high statistical accuracies of recent EAS experiments 
\cite{Tibet,EAS-TOP,CASA,KASCADE1} have confirmed the presence of
a bend in the all-particle primary
energy spectrum at around $3\cdot10^3$ TeV (called the "knee")
from an overall spectrum $\propto E^{-2.7}$ below the knee 
to $\propto E^{-3.1}$
beyond the knee,
and a change in composition toward heavier species with increasing
energy in the $10^3-10^5$ TeV region.
However, separating the primary energy spectra of elemental groups
remains difficult, both due to uncertainties in the interaction
model and the uncertainties associated with the solutions to the
EAS inverse problem \cite{MACRO,KASCADE2}.

One of the most studied class of models for the origin of cosmic rays
in this energy region, which assumes that supernova remnants are their
main source, predicts rigidity-dependent primary energy spectra in the
knee region (\cite{Stanev,Hillas1} and references therein).  Other
astrophysical models for the origin of the knee, such as Galactic
propagation effects \cite{Peters,Hillas2} also predict rigidity-dependent spectra.  
Such  energy spectra of primary nuclei with rigidity-dependent knees
can approximately describe 
the observed EAS size spectra in the $10^3-10^5$ TeV energy region
in the framework of conventional interaction models
\cite{CASA1,Samo,Ter-Antonyan3,EAS-TOP1,Ter-Antonyan5}.
However, an alternative class of models predicts mass-dependent knees
(see \cite{Hoerandel} and references therein for a recent review of
models of the origin of the knee).
In the present analysis, we will assume a rigidity-dependent knee;
the appropriateness of this hypothesis will be briefly examined in
the discussion section.

The GAMMA facility (Fig.~1) was designed at the beginning of 
the 1990's in
the framework of the ANI experiment \cite{ANI} and the first 
results of EAS investigations were presented in 
\cite{GAMMA00,GAMMA01,GAMMA02,GAMMA03}.
The main characteristic features of the GAMMA experiment are the 
mountain location, the symmetric location of the EAS detectors,
and the underground muon scintillation carpet which detects the
EAS muon component with energy $E_{\mu}>5$ GeV.

Here, a description of the GAMMA facility, the main results 
of investigation during 2002-2004 \cite{GAMMA02,GAMMA03,Martir}
and evaluations of primary energy spectra in the knee region 
are presented in comparison with the 
corresponding simulated data in the
framework of the SIBYLL \cite{SIBYLL} and QGSJET \cite{QGSJET}
interaction models. Preliminary results have already been   
presented in \cite{GAMMA05,ICRC29a,ICRC29b}.

\section{GAMMA experiment} 
The GAMMA installation \cite{GAMMA00,GAMMA01,GAMMA02,GAMMA03,Martir}
is a ground-based array of 33 surface 
particle detection stations and 150 underground muon detectors,
located on the south side of Mount Aragats in Armenia.
The elevation of the GAMMA facility is 3200 m above sea level,
which corresponds to 700 g/cm$^2$ of atmospheric depth. A
diagrammatic layout is shown in Fig.~1.\\
\begin{figure} 
\begin{center}
\includegraphics[width=14.714cm,height=6cm]{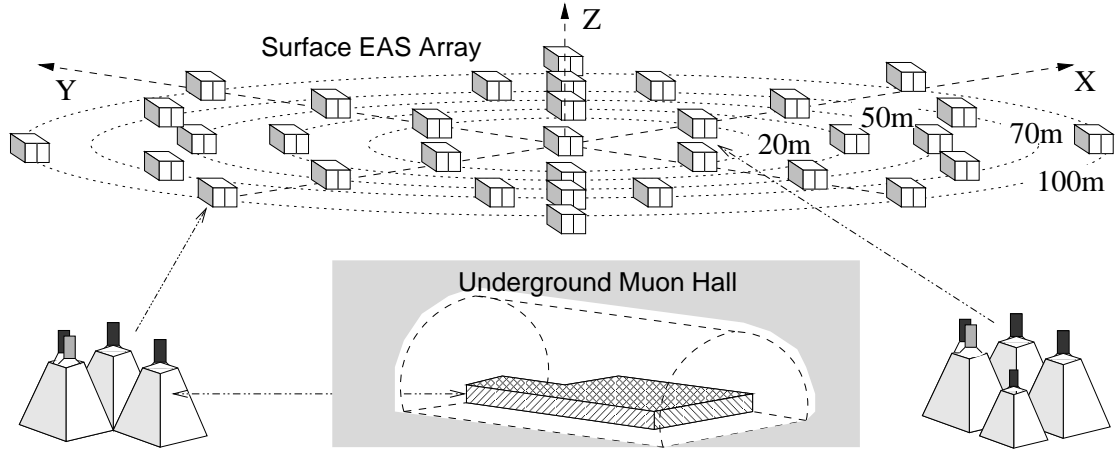}
\end{center}
\caption{Diagrammatic layout of the GAMMA facility.}
\vspace{10mm}
\end{figure}
The surface stations of the EAS array are located on 5
concentric
circles of radii $\sim$20, 28, 50, 70 and 100 m, and each station contains
3 square plastic scintillation detectors with the following dimensions:
$1\times1\times0.05$ m$^3$. Each of the central 9 stations contains an 
additional
(4th) small scintillator with dimensions $0.3\times0.3\times0.05$ m$^3$ 
(Fig.~1) for high
particle density ($\gg10^2$ particles/m$^2$) measurements.\\
A photomultiplier tube is positioned on the top of the aluminum
casing covering each scintillator. One of the three detectors of each 
station is examined by two photomultipliers, one of which is designed 
for fast timing measurements.\\
150 underground muon detectors (muon carpet) are
compactly arranged in the underground
hall under 2.3 kg/cm$^2$ of concrete and rock. The scintillator 
dimensions, casings and 
photomultipliers are the same as in the EAS
surface detectors.
                                                   
\subsection{Detector system and triggering} 
The output voltage of each photomultiplier is converted into 
a pulse burst by a logarithmic ADC and transmitted to a CAMAC 
array, where
the corresponding electronic counters produce a digital number
(``code'') of pulses in the burst. Four inner (``trigger'') stations 
at a radius of 20m are monitored by a coincidence circuit. 
If at least two scintillators of each
trigger station each detect more than 3 particles, the information 
from
all detectors are then recorded along with the time between the 
master trigger pulse and the pulses from all fast-timing detectors. 
The given trigger condition provides EAS detections with an EAS size
threshold $N_{ch}>(0.5-1)\cdot10^5$ for a location of the EAS core
within the $R<50$ m circle. The shower size thresholds
for $100\%$ shower detection efficiency are equal to
$N_{ch}=3\cdot10^5$ and $N_{ch}=5\cdot10^5$ for EAS core locations
within $R<25$ m and $R<50$ m respectively \cite{GAMMA00}.\\
Before being placed on the scintillator
casing, all photomultipliers were tested by a test bench using a 
luminodiode method where
the corresponding parameters of the logarithmic ADC and the upper 
limit ($(0.5-1)\cdot10^4$) \cite{PION} of the particle density 
measurement ranges were determined for each detector.
The number of charged particles ($n_i$) passing through the i-th 
scintillator is calculated using a logarithmic 
transformation: $\ln n_i=(C-C_0)/d$ \cite{PION}, where 
the scale parameter $d\simeq(9-10)\pm0.35$ is determined
for each detector
by the test bench, $0\leq C\leq2^{7}-1$ is the output digital 
code from the CAMAC array corresponding 
to the energy deposit of $n$ charged particles into the 
scintillator, and $C_0\simeq(5-6)\pm0.25$ is equal to 
the mode of the background single particle digital code spectra
(Section 2.4).
The time delay is estimated by the pair-delay method 
\cite{Ter-Antonyan1} to give a time resolution of about $4-5$ ns.    

\subsection{Reconstruction of EAS parameters} 
The EAS zenith angle ($\theta$) is estimated
on the basis of the shower front arrival times measured
by the 33 fast-timing surface detectors,
applying a maximum likelihood method and the flat-front approach
\cite{Ter-Antonyan1,MAKET}.
The corresponding uncertainty was tested by 
Monte Carlo simulations and is equal to 
$\sigma(\theta)\simeq1.5^\circ$ \cite{GAMMA00}.
The reconstruction of the EAS size ($N_{ch}$), shower age ($s$)
and core coordinates ($x_0,y_0$) is performed based on the
Nishimura-Kamata-Greisen (NKG) approximation to the measured charged 
particle densities
($\{n_i\},i=1,\dots,m$), using $\chi^2$ minimization
to estimate $x_0,y_0$ and a maximum
likelihood method to estimate $N_{ch}$, taking
into account the measurement errors. Gamma-quanta conversions
in the scintillator and housing were taken into account
in the estimates of $N_{ch}$ (Section 2.3).\\ 
The logarithmic transformation $L(n_i)=\ln n_i-(1/m)\sum\ln
n_i$ for $n_i\neq0$ enables an analytical solution
for the EAS age parameter ($s$) using $\chi^2$ minimization
\cite{MAKET,Ter-Antonyan2}.
Unbiased ($<5\%$) estimations of $N_{ch},s,x_0$ and $y_0$ shower
parameters are obtained for $N_{ch}>5\cdot10^5$, 
$0.3<s<1.6$, $\theta<30^\circ$ and distances $R<25$ m from the shower
core to the center of the EAS array. The shower age parameter ($s$)
is estimated from the surface scintillators located inside
a $7$ m $<R_i< 80$ m ring area around the shower core (Section 2.3).\\
The EAS detection efficiency ($P_d$) and corresponding accuracies are 
derived from mimic shower simulations taking into account the EAS
fluctuations and measurement errors (Section 2.4)
and are equal to: $P_d=100\%$, 
$\Delta N_{ch}/N_{ch}\simeq0.1$, 
$\Delta s\simeq0.05$, 
$\Delta x$ and $\Delta y\simeq0.5-1$ m. 
These results were also checked with CORSIKA \cite{CORSIKA} 
simulated EAS (Section 2.3) and depend slightly on shower core
location for $R < 50$ m.\\
The reconstruction of the total number of EAS muons ($N_{\mu}$) 
from the detected muon densities ($\{n_{\mu,j}\}, j=1,\dots,150$)
in the underground muon hall is carried out by restricting
the distance to $R_{\mu}<50$ m from the shower core
(the so-called EAS ``truncated'' muon size \cite{GAMMA00,KASCADE01}) 
and using 
the approximation to the muon lateral distribution function
\cite{GAMMA00,Hillas}: 
$\rho_{\mu}(r)=cN_{\mu}(R_{\mu}<50{\rm m})\exp{(-r/r_0)}/(r/r_0)^{0.7}$,
where $r_0=80$ m \cite{TS} and $c=1/2\pi\int_{0}^{50} \rho(r)rdr$. 
The EAS truncated muon size $N_{\mu}(R_{\mu}<50 {\rm m})$ is estimated
at known (from the EAS surface array) shower core coordinates in the
underground muon hall. Unbiased estimations for muon size
are obtained for $N_{\mu}>10^3$ using a maximum likelihood method
and assuming Poisson fluctuations in the detected muon numbers.
The reconstruction accuracies of the truncated muon shower sizes
are equal to $\Delta N_{\mu}/N_{\mu}\simeq0.2-0.35$
at $N_{\mu}\simeq10^5-10^3$ respectively.\\
It should be noted that the detected muons in the 
underground hall are always accompanied by the electron-positron 
equilibrium spectrum which is produced when muons pass through the 
matter (2300 g/cm$^2$) over the scintillation carpet; this
is taken into account in our results (Section 3.2). 

\subsection{Detector response} 
The GAMMA detector response taking into account the EAS
$\gamma$-quanta contribution was computed by 
EAS simulations using the CORSIKA 6.031 code \cite{CORSIKA}
(NKG and EGS modes, GHEISHA2002)
with the QGSJET01 \cite{QGSJET} and SIBYLL 2.1 \cite{SIBYLL}
interaction models for 4 types ($A\equiv H,He,O,Fe$) of primary
nuclei. Each EAS particle ($\gamma,e,\mu,h$) obtained from 
CORSIKA (EGS mode) at the observation level was examined by
passing through the steel casing (1.5 mm) of the detector station 
and then through the corresponding scintillator. 
The pair production and Compton scattering 
processes were additionally simulated in the case of 
EAS $\gamma$-quanta.
The resulting energy deposit in the scintillator 
was converted to an 
ADC code and inverse-decoded into a number of ``detected'' charged 
particles taking into account all uncertainties of the ADC parameters 
($C_0,d$) and fluctuations in the light collected by 
the photomultipliers ($\sigma_l\simeq0.25/\sqrt{n}$).\\
Using the simulation scenario above, 200 EAS events with shower
size threshold $N_{ch}>5\cdot10^5$ were 
simulated with CORSIKA simultaneously in the
EGS and NKG modes for each of the $A\equiv H,He,O$ and $Fe$ primary nuclei,
with logarithm-uniform energy spectra in the $10^3-10^5$ TeV 
energy range. The computation of the charged particle 
densities in surface detectors in the NKG mode was 
performed by applying two-dimensional  
interpolations of the corresponding shower electron (and positron)
density matrix from CORSIKA \cite{CORSIKA}, along with
the individual EAS muons and hadrons.\\
\begin{figure}[h]  
\begin{center}
\includegraphics[width=8cm,height=8cm]{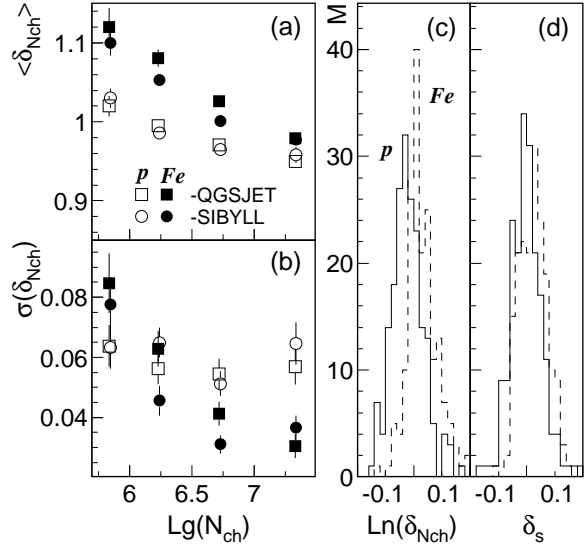}
\end{center}
\caption{Biases (a) and standard deviations (b) versus
EAS size, and distributions of biases in shower size (c)
and shower age parameter (d) for the SIBYLL 
(circle symbols, lines)
and QGSJET (square symbols) interaction models
for primary $H$ (empty symbols, solid lines) and $Fe$ 
(filled symbols, dashed lines) nuclei.}
\vspace{10mm}
\end{figure}
A $\sim5\%$ agreement between the EGS (including the EAS
$\gamma$-quanta contribution) and NKG
simulated EAS data was attained for an 
$E_e\simeq1$ MeV kinetic energy
threshold of shower electrons (and positrons) in the NKG mode,
considering only the 7m $<R_i<80$m ring area used
in the determination of the shower age parameter.  Thus
the underestimation of the EAS particle density due
to the threshold of the detected energy deposit ($E_d\simeq8$ MeV
\cite{GAMMA00,GAMMA05}) 
in the scintillators is compensated by the EAS $\gamma$-quanta 
contribution.\\
The corresponding biases 
\begin{displaymath}
\delta_{N_{ch}}(A,N_{ch})\equiv
\frac{N_{ch}(E_e=1{\rm\,MeV},NKG)}{N_{ch}(E_d,\gamma,EGS)}
\end{displaymath}
and standard deviations $\sigma(\delta_{N_{ch}})$ versus 
the reconstructed EAS size ($N_{ch}$) are shown in Fig.~2 (a) and (b)
respectively, for the SIBYLL (circle symbols) and QGSJET 
(square symbols) interaction models and for primary $H$ 
(empty symbols) and $Fe$ nuclei (filled symbols).
The distributions of the biases
in reconstructed EAS sizes ($\delta_{N_{ch}}$)
and shower age parameters 
\begin{displaymath}
\delta_s(A)\equiv s(E_e=1{\rm\,MeV},NKG)-s(E_d,\gamma,EGS) 
\end{displaymath}
are shown in Fig.~2 (c) and (d) respectively, for a shower size
threshold $N_{ch}>5\cdot10^5$, the SIBYLL interaction model, and
primary $H$ (solid lines) and $Fe$ (dashed lines) nuclei.\\
The observed ($\sim 5\%$) biases in $\delta_{N_{ch}}$ (Fig.~2a)
for the 4 kinds of primary nuclei depend only weakly
on the interaction model ($\leq 5\%$) and zenith angles 
($\leq3\%$ for $\theta <30^\circ$), and the biases in
age parameter $\delta_s$ can be considered negligible.
The NKG-mode simulated sizes were divided
by the estimated biases $\delta_{N_{ch}}(A, N_{ch})$
in the reconstruction of the primary energy spectra (Section 3.1).

\subsection{Measurement errors and density spectra} 
The close disposition of the $k=1,2,3$ scintillators in each of the 
($i$-th) detector stations of the GAMMA surface array  
enables a calibration of the measurement error
using the detected EAS data. 
The measured and simulated particle
density discrepancies $(n_{k}-\rho)/\rho$ versus 
the average value $\rho=(1/3)\sum{n_{k}}$ for distances
$R_i>10$ m from the shower core are shown in Fig.~3 
(circle symbols), and are completely determined 
by Poisson fluctuations (at $R_i\gg1$ m ) and the measurement errors.
\begin{figure}[h]   
\begin{center}
\includegraphics[width=7cm,height=7cm]{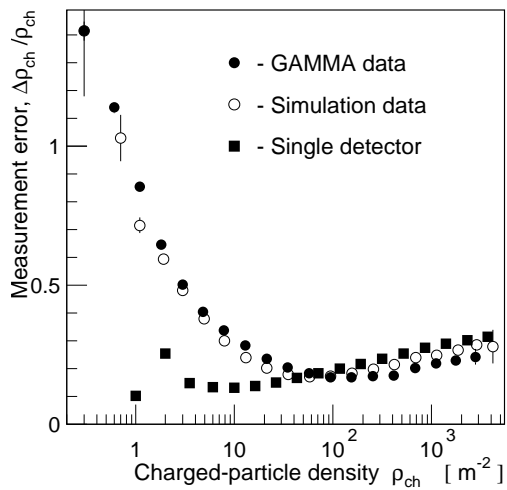}
\end{center}
\caption{Particle density discrepancies
(circle symbols) and measurement error for a single detector 
(square symbols) versus charged-particle density.}
\vspace{10mm}
\end{figure}
The agreement between the measured and simulated dependences enables
the extraction of the actual measurement errors of the GAMMA 
detectors.  The corresponding results, obtained from the
simulations without Poisson fluctuations, are shown in Fig.~3
(square symbols).\\
The background omni-directional single particle spectra (in
units of ADC code) detected by GAMMA  
surface scintillators in 78 s of operation time are shown 
in Fig.~4 (dotted lines).
The background single particle spectra detected by underground 
\begin{figure}[h]   
\begin{center}
\includegraphics[width=7cm,height=7cm]{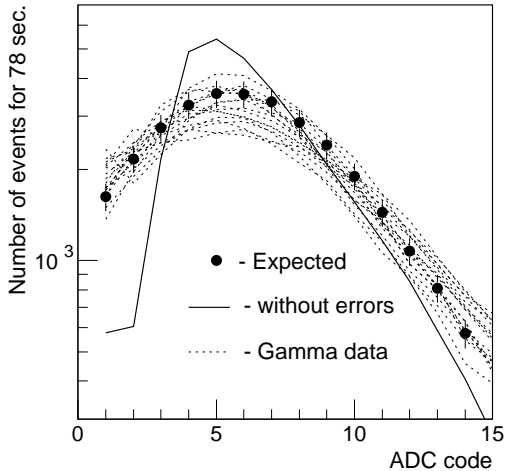}
\end{center}
\caption{Background single particle spectra of 15 surface detectors
(dotted lines). The symbols (solid line) are the expected spectra 
taking into account (without) measurement errors.}
\vspace{10mm}
\end{figure}
muon scintillators have the same shape but about 10 times lower
intensities. These spectra (pulse height distributions) along with 
the known zenith angle distributions and composition
($\sim40\%e,50\%\mu$) of the background charged particles 
at the observation level \cite{2TER} are used for the operative  
determination of the ADC parameters ($C_0$) for each experimental run.
The symbols and solid lines in 
Fig.~4 display the corresponding expected spectra 
obtained by CORSIKA (EGS) simulation, without errors (solid line)
and taking into account the measurement errors (symbols)
respectively. The minimal primary energy in the
simulation of the background particle spectra was determined
by the 7.6 GV geomagnetic rigidity cutoff in Armenia.\\
Because the effective primary energies responsible for the 
single particle spectra at the observation level of 700 g/cm$^2$ 
are around $100$ GeV, and this energy range
has been studied by direct measurements in balloon and satellite
experiments, the primary energy spectra and 
elemental composition in the Monte Carlo simulation were taken from 
power-law approximations to the direct measurement data
\cite{Wiebel}. It should be noted that the expected single particle spectra
at these energies are practically the same for the QGSJET and SIBYLL
interaction models, because most of the interactions occur in
the energy range where accelerator data are used.\\
Fig.~5 (symbols) displays the EAS charged particle density
spectra measured by the surface
detectors (left panel) and underground muon detectors (right panel) 
at $R_i<50$ m with different EAS size thresholds: 
$N_{ch}>5\cdot10^5$, $N_{ch}>10^7$ (and additionally 
$N_{ch}>2\cdot10^6$ for the muon density spectra).
\begin{figure}   
\begin{center}
\includegraphics[width=13cm,height=7cm]{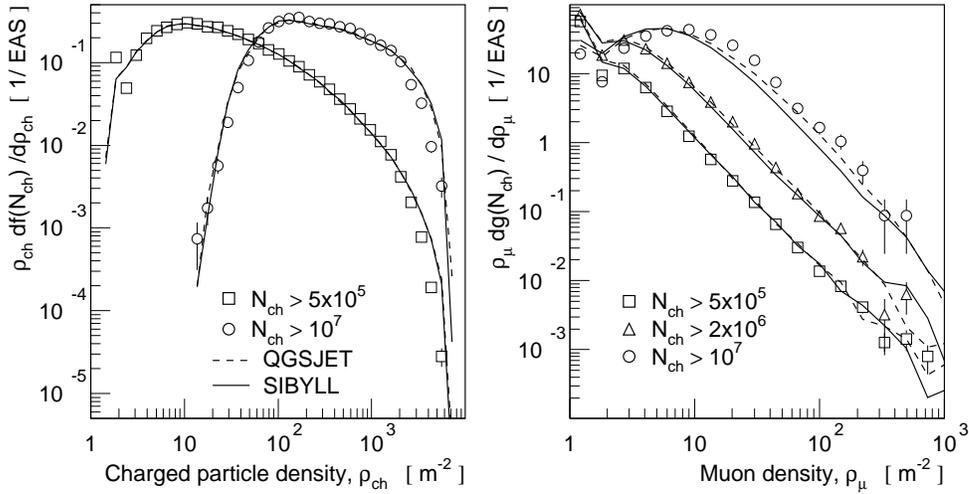}
\end{center}
\caption{Detected (symbols) and expected (lines) 
particle density spectra measured by the surface scintillators 
(left panel) and underground muon scintillators (right panel).}
\vspace{10mm}
\end{figure}
The showers were selected with $\theta<30^\circ$ and shower core
location in the $R<25$ m range from the center of the GAMMA facility 
(Fig.~1). The corresponding expected spectra 
(lines) for different interaction
models are also shown in Fig.~5. The primary energy spectra
and elemental composition for these
simulations were those obtained in the combined approximation
solution to the EAS inverse problem 
(Section 3.3). There is 
good agreement between the expected and observed data for the surface
array over the full measurement range (about four orders of
magnitude).  However, agreement of the detected 
muon density spectra with the expected ones 
is attained only in the $N_{ch}<10^7$ range. The 
observed discrepancies for the muon density spectra at $N_{ch}>10^7$ 
are unaccounted for at present, and will require subsequent
investigations.

\subsection{EAS data set} 
The data set analyzed in this paper was obtained
over $6.19\cdot10^7$ s of operating live time of the GAMMA
facility, from 2002 to 2004.
Showers were selected for analysis with the following
criteria: $N_{ch}>5\cdot10^5$, $R<25$ m, 
$\theta<30^\circ$, $0.3<s<1.6$, $\chi^2(N_{ch})/m<3$ and  
$\chi^2(N_{\mu})/m<3$ (where $m$ is the number of scintillators
with non-zero signal),
yielding a total data set of $1.9\cdot10^5$ selected showers.
\begin{figure}   
\begin{center}
\includegraphics[width=7cm,height=7cm]{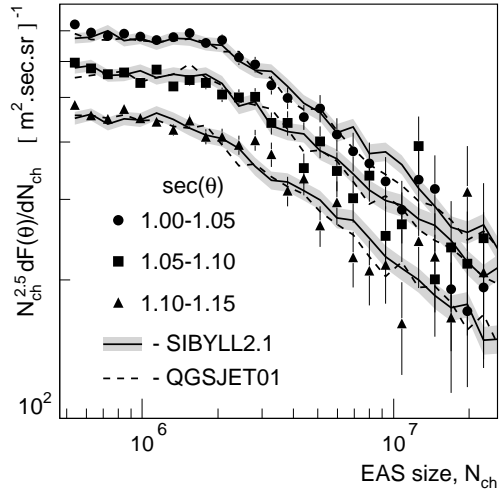}
\end{center}
\caption{EAS size spectra for 3 zenith
angle intervals (symbols) and corresponding expected spectra
according to the SIBYLL (solid lines) and QGSJET (dashed lines)
interaction models.} 
\vspace{10mm}
\end{figure}
\begin{figure}   
\begin{center}
\includegraphics[width=7cm,height=7cm]{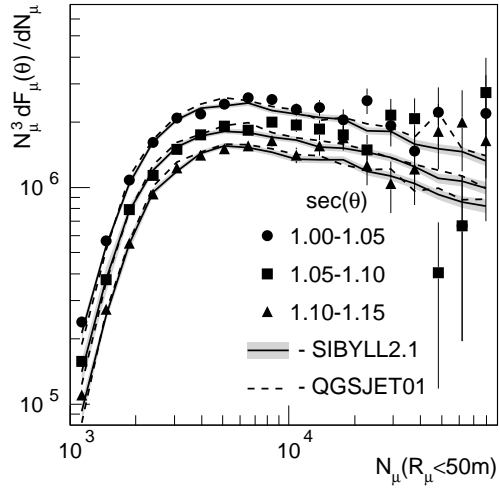}
\end{center}
\caption{Normalized EAS truncated muon size spectra for 3 zenith
angle intervals (symbols) and corresponding 
expected spectra for the SIBYLL (solid lines) and QGSJET 
(dashed lines) interaction models.}
\vspace{10mm}
\end{figure}
\begin{figure}   
\begin{center}
\includegraphics[width=7cm,height=7cm]{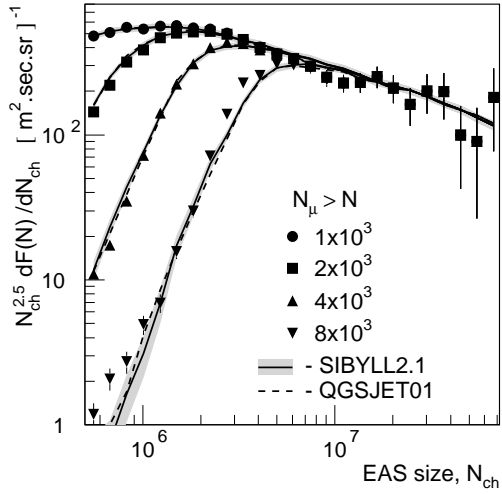}
\end{center}
\caption{EAS size spectra (symbols)
for different truncated muon size thresholds ($N_{\mu}$)
and corresponding expected spectra according to the SIBYLL
(solid lines) and QGSJET (dashed lines) interaction models.}
\vspace{10mm}
\end{figure}
\begin{figure}   
\begin{center}
\includegraphics[width=7cm,height=7cm]{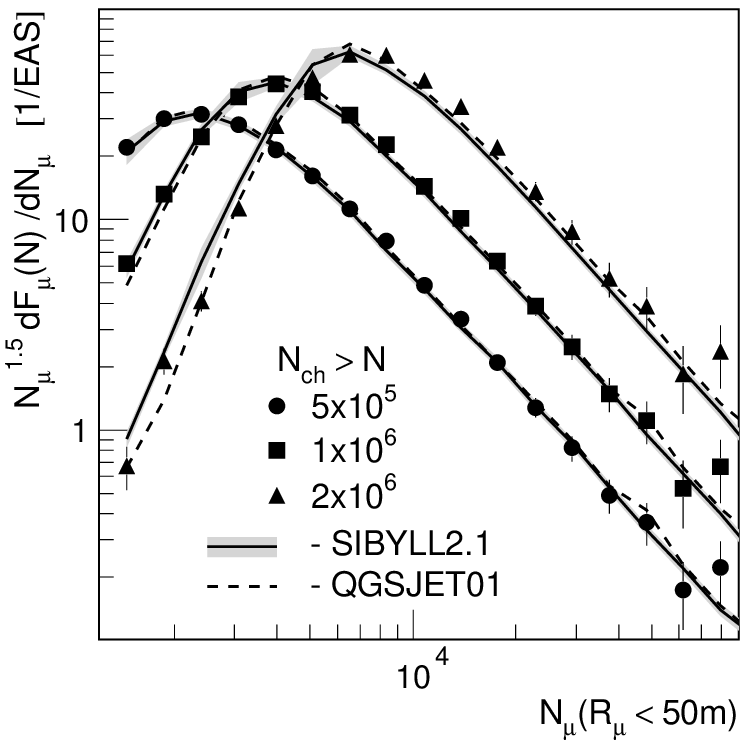}
\end{center}
\caption{Normalized EAS truncated muon size spectra (symbols) for different
shower size thresholds ($N_{ch}$) and corresponding expected spectra 
according to the SIBYLL (solid lines) and 
QGSJET (dashed lines) interaction models.}
\vspace{10mm}
\end{figure}
The selected measurement range 
provided $100\%$ EAS detection efficiency (Section 2.2)
and similar conditions for the reconstruction of the shower 
lateral distribution functions.\\
The measured variable distributions used in the combined
approximation approach to the EAS inverse problem (Section 3.3)
are shown in Figs.~6--11 (symbols).
All lines and shaded areas in these figures correspond
to the expected spectra computed on the basis of the EAS inverse problem
solution in the framework of the SIBYLL and QGSJET  
interaction models. 
These expected (forward folded) spectra are computed by
Monte-Carlo integration (Section 3.1) using the simulated EAS
database, which results in the statistical fluctuations evident
in many of these predicted spectra.\\
\begin{figure}   
\begin{center}
\includegraphics[width=7cm,height=7cm]{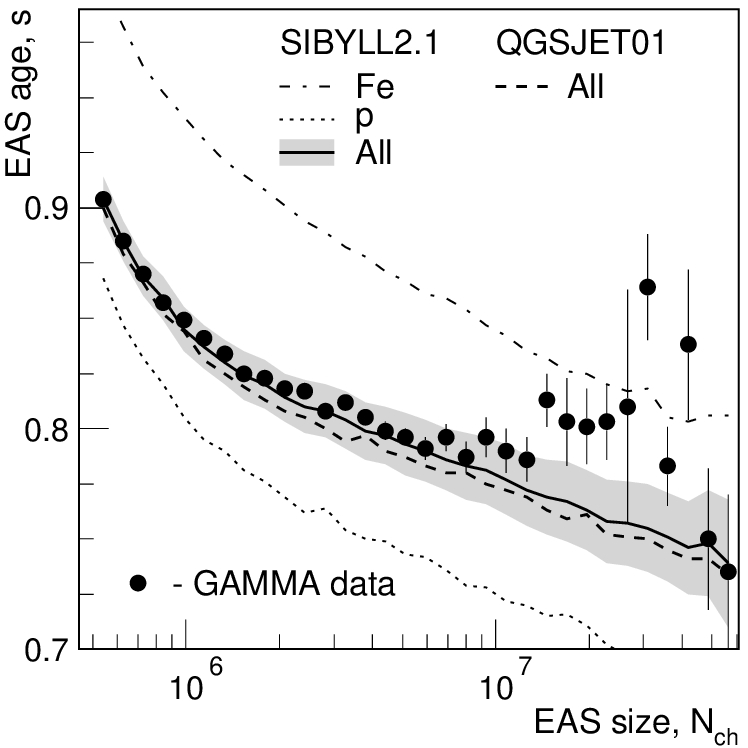}
\end{center}
\caption{Dependence of the average EAS age parameter on EAS size
(symbols) along with expected values 
for the SIBYLL (solid line) and QGSJET (dashed line) interaction models. 
The dotted and dash-dotted lines correspond to expected 
values for primary hydrogen and iron nuclei, respectively, for the 
SIBYLL interaction model.}
\vspace{10mm}
\end{figure}
\begin{figure}   
\begin{center}
\includegraphics[width=7cm,height=7cm]{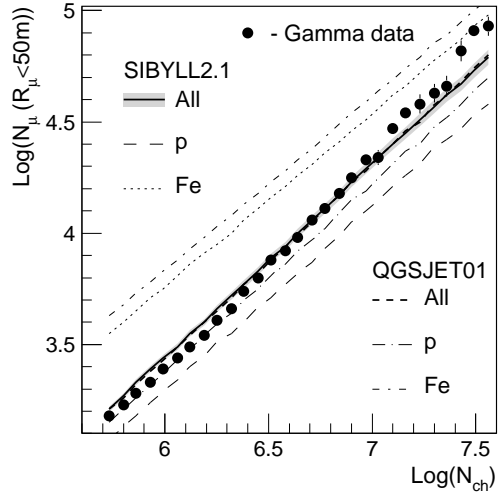}
\end{center}
\caption{Average EAS truncated muon size
$N_{\mu}$ 
versus EAS size $N_{ch}$
(symbols). The lines and shaded area are the expected dependences 
for the SIBYLL and QGSJET interaction models and primary $p$, $Fe$ and 
mixed (Table~1) compositions.}
\vspace{10mm}
\end{figure}
The EAS size spectra ($N^{2.5}_{ch}\cdot \d F(\theta)/\d N_{ch}$) for
three zenith angle intervals are shown in Fig.~6.
The EAS truncated muon size spectra  
in the same zenith angle intervals are shown in Fig.~7; 
these spectra are normalized per unit shower with
$N_{ch}>5\cdot10^5$ and $\theta<30^\circ$.
The EAS size spectra for $\theta<30^\circ$ and 
different thresholds in EAS truncated muon size are shown in 
Fig.~8.
The normalized EAS truncated muon size spectra
for different EAS size thresholds are shown in Fig.~9. 
Fig.~10 displays the dependence of the average EAS age parameter   
on EAS size $s(N_{ch})$.  
Fig.~11 shows
the observed $N_{\mu}(N_{ch})$ dependence and the corresponding
expected values for primary proton, iron and mixed
($p,He,O,Fe$, Section~3.3) compositions computed 
in the framework of the SIBYLL and QGSJET interaction models.

\section{EAS inverse problem and primary energy spectra} 
\subsection{Key assumptions } 
The observed spectra $F(\mathbf{q})$ of the measured EAS parameters 
$\mathbf{q}=(N_{ch}, N_\mu, s)$ 
result from convolutions of the 
energy spectra $I_A(E)$ of primary nuclei ($A\equiv H,He,\dots$
at least up to $Ni$) with the probability density distributions
$W_A(E,\mathbf{q})$ 
\cite{Ter-Antonyan3,KASCADE01,Ralph}:
\begin{equation}
F(\mathbf{q})=
\sum_{A}\int_{E}W_A(E,\mathbf{q})I_{A}(E)
\d E\;.
\end{equation}
The functions $W_A(E,\mathbf{q})$ are derived using
a model of the EAS development in the atmosphere and
convolution of the resulting shower spectra at the observation
level with the corresponding response functions 
\cite{KASCADE2,GAMMA05}.\\
The integral equation (1) defines the EAS inverse
problem, namely the evaluation of the primary energy 
spectra $I_{A}(E)$ on the basis of the measured distributions
$F(\mathbf{q}_i)$ (in $i=1,\dots,V$ discrete bins)
and the known kernel functions $W_A(E,\mathbf{q}_i)$ 
\cite{KASCADE2,GAMMA05,Ralph}.
The multidimensional kernel functions $W_A(E,\mathbf{q})$ can be
computed using interpolations \cite{Ter-Antonyan3} or approximations
\cite{KASCADE2} to the corresponding spectra, which  
are previously obtained by CORSIKA EAS simulations 
in the framework of a given interaction model, for 
different groups of primary nuclei and a set of primary energies
and zenith angles.

In the present work,
the computations of the expected shower spectra (forward folding) 
from (1) for given primary energy spectra $I_{A}(E)$ are
performed by Monte Carlo integration \cite{M-C,Num-Rec},
using an arbitrary positive weight function $I_0(A,E)$ determined
in the same domain as the primary spectra $I_{A}(E)$ and normalized
such that $\int W_A(E)I_0(A,E)\d E=1$.\\
Multiplying and dividing the integrand in (1) by $I_0(A,E)$, 
expression (1) is converted to the form: 
\begin{equation}
F(\mathbf{q}_i)=
\sum_{A}
\Big<
\frac{I_{A}(E)}{I_0(A,E)}
\Big>_{\mathbf{q}_j\in \mathbf{q}_i} \;.
\end{equation}
The averaging in (2) is performed 
over random $E_j$ ($j=1,\dots,\mathcal{N_A}$) distributed with
a probability density function 
$I_0(A,E)W_A(E\mid\mathbf{q}_j\in\mathbf{q}_i)$, with
shower parameters $\mathbf{q}_j$ within the given $\mathbf{q}_i$ bin.
The reconstructed shower parameters $\mathbf{q}_j(A,E_j)$
are obtained by EAS simulations in the framework of a 
given interaction model, taking into account the 
corresponding response functions $<\delta_{N_{ch}}(A,N_{ch})>$ 
(Section 2.3).\\
As a weight function we chose the power law spectrum 
$I_0(A,E)\propto E^{-1.5}$ which provides an accuracy for integration 
$\Delta F/F\simeq 1/\sqrt{\mathcal{N}}$ and relatively
small statistical errors for the simulated EAS samples both within 
and especially beyond the knee region.
The accuracy of Monte-Carlo integration with this weight
function was checked using power-law spectra
$f(x)\propto x^{-\gamma}$ with $\gamma=2.5-3.3$ and
log-normal distributions $W(x,y)$, and found to be adequate
for our purposes.

In order to evaluate the primary energy spectra on the basis of 
the EAS data set we regularized the integral equation (1)
using a parametrization method \cite{Ter-Antonyan3,Ter-Antonyan5}.
The solutions for the primary energy spectra in (1) were sought 
based on a broken power-law function 
with a ``knee'' at the rigidity-dependent 
energy $E_k(A)=E_R\cdot Z$, and the same spectral indices
for all species of primary nuclei ($A\equiv p,He,O,Fe$), 
$\gamma_1$ below and $\gamma_2$
above the knee respectively:
\begin{equation}
\frac{\d I_A}{\d E}=\Phi_A \,
\left(\frac{E_k}{1\,{\rm TeV}}\right)^{-\gamma_1}
\left(\frac{E}{E_k}\right)^{-\gamma} \: ,
\end{equation}
where $\gamma=\gamma_1$ for $E\leq E_k(A)$,
$\gamma=\gamma_2$ for $E>E_k(A)$, $E_R$ is the particle's 
magnetic rigidity
and $Z$ the charge of nucleus $A$.\\
The integral equation (1) is thereby transformed into a parametric
equation with the unknown spectral parameters
$\Phi_A,E_R,\gamma_1$ and $\gamma_2$,
which are evaluated by minimization of the $\chi^2$ function:
\begin{equation}
\chi^2=\sum_{u=1}^{U}\sum_{i=1}^{V_u}
\frac{(\zeta_{u,i}-\xi_{u,i})^2}
     {\sigma^2(\zeta_{u,i}) +\sigma^2(\xi_{u,i})} \; ,
\end{equation}
where $U$ is  the number of examined functions 
$\zeta_{u,i}\equiv F_u({q}_{u,i})$  
obtained from the experimental data with statistical accuracies 
$\sigma(\zeta_{u,i})$ in $i=1,\dots,V_u$ bins, and
$\xi_{u,i}$ and $\sigma(\xi_{u,i})$ are the corresponding 
expected (forward folded) values from (2) and their
(statistical) uncertainties.

\subsection{Simulated EAS database } 
EAS simulations for the evaluation of 
the primary energy spectra using the GAMMA facility EAS data 
were carried out for $\mathcal{N}_A\equiv$
$10^5$ primary $H$, $7.1\cdot10^4$ $He$,
$4.6\cdot10^4$ $O$ and $4.8\cdot10^4$ $Fe$ nuclei 
using the CORSIKA NKG mode and the SIBYLL interaction model.
The corresponding statistics for the 
QGSJET interaction model were: 
$10^5$, $6\cdot10^4$, $4.4\cdot10^4$ and $4\cdot10^4$.\\
The energy thresholds of the primary nuclei were the same for both
interaction models and were set at $E_{A,\min}\equiv0.5,0.7,1$ and 
$1.2$ PeV for $H$, $He$, $O$ and $Fe$ respectively, and the
upper energy limit was set at $E_{\max}=5\cdot10^3$ PeV.
The simulated energies were distributed following a weight
function $I_0(A,E) \propto E^{-1.5}$, as explained above.
The simulated showers had core coordinates distributed
uniformly within a radius $R < 25$ m, and zenith angles
$\theta < 30^\circ$.
This ignores the effect of showers with true core coordinates
outside the selection radius which have reconstructed coordinates
with $R < 25$~m; due to the good core reconstruction
accuracy of $0.5 - 1$~m (Section 2.2), this effect may be
neglected for our purposes.\\
All the EAS muons with energies of $E_{\mu}>4$ GeV at the GAMMA observation
level were passed through the  
2.3 kg/cm$^2$ of rock to the muon scintillation carpet (the underground 
muon hall). The fluctuations in the muon ionization losses, and the
electron (and positron)
accompaniment due to the muon electromagnetic and photonuclear  
interactions in the rock were taken into account, using 
the approximation of an equilibrium accompanying 
charged particle spectrum obtained from preliminary
simulations with the FLUKA code \cite{FLUKA} in the $0.005-20$ TeV
muon energy range.  The resulting
charged particle accompaniment per EAS muon in the underground hall
is equal to $0.06\pm0.01$ ($100\%e$) and $11.0\pm1.5$
($98.5\%e,1.4\%h,0.04\%\mu$) at muon energies $0.01$ 
TeV and $10$ TeV respectively.\\
The total number of simulated EAS in the database were 
$\mathcal{N}=\sum\mathcal{N}_A\simeq2.65\cdot10^5$ EAS for the SIBYLL 
and $\mathcal{N}\simeq2.44\cdot10^5$ EAS for the QGSJET 
model.

\subsection{Combined approximations to the EAS data} 
Using the aforementioned formalism (Section 3.1), the $U=6$ examined
functions shown in Figs.~6--11 and the corresponding EAS data set, 
the unknown spectral parameters $\Phi_A,E_R,\gamma_1$ and
$\gamma_2$ of parametrization (3) were derived by minimization of the
$\chi^2$ (4) and forward folding (2), with a number of degrees of freedom  
$n_{d.f.}=\sum_{1}^{6}{V_u}-p-1\simeq350$, where $p=7$ is the number
of adjustable parameters.\\ 
The values of the spectral parameters (3) derived from the solution
of the parameterized equation (1) are presented in Table~1
for the SIBYLL and QGSJET interaction models.
\begin{table}  
\caption{\label{tab:table1}
Parameters of the primary energy spectra (3) 
from combined approximations to the EAS data. 
The scale factors $\Phi_A$
and particle rigidity $E_R$
respectively have units of (m$^2\cdot$ s $\cdot$ sr $\cdot$ TeV)$^{-1}$
and TV.}
\begin{center}
\begin{tabular}{lcc}
\hline
Parameters & SIBYLL&QGSJET\\
\hline
$\Phi_  H$ & $0.095\pm0.008$ & $0.165\pm0.005$\\
$\Phi_{He}$& $0.100\pm0.012$ & $0.020\pm0.008$\\
$\Phi_O   $& $0.034\pm0.007$ & $0.008\pm0.004$\\
$\Phi_{Fe}$& $0.024\pm0.004$ & $0.013\pm0.005$\\
$E_R      $& $2500\pm200$    & $3200\pm150$   \\
$\gamma_1$ & $2.68\pm0.015$ & $2.66\pm0.010$\\
$\gamma_2$ & $3.19\pm0.03$   & $3.11\pm0.02$\\
$\chi^2/n_{d.f.}$  & $2.0$ &$1.5$\\
\hline
\end{tabular}
\end{center}
\end{table}
\begin{figure}   
\begin{center}
\includegraphics[width=13cm,height=8cm]{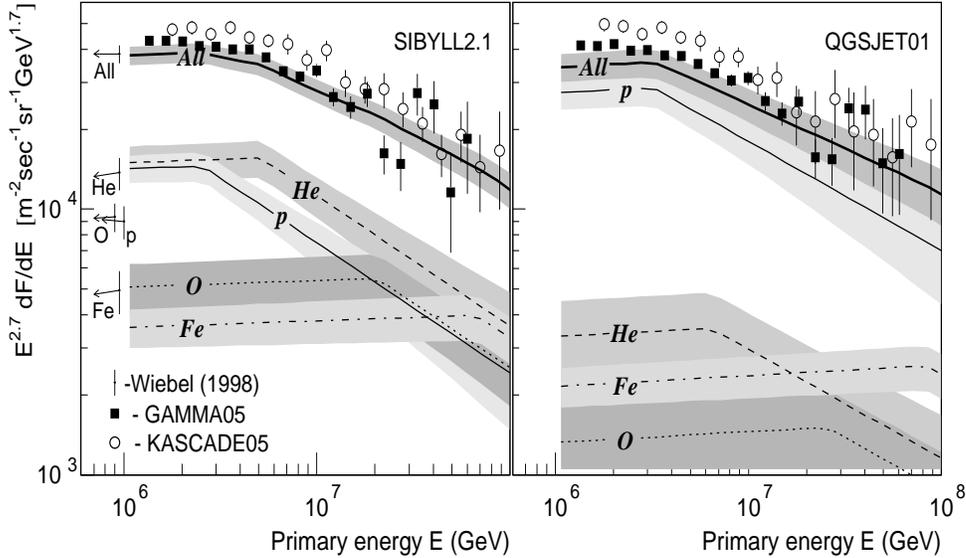}
\end{center}
\caption{Energy spectra and abundances of the primary nuclei groups
(lines and shaded areas) for the SIBYLL (left panel) and
QGSJET (right panel) interaction models.  All-particle spectra
from GAMMA \cite{ICRC29b} and KASCADE \cite{KASCADE2} are shown
as symbols. Vertical bars show the 
extrapolations of balloon and satellite data \cite{Wiebel}.}
\vspace{10mm}
\end{figure}
The primary energy spectra obtained for $p, He, O,$ and $Fe$ nuclei,
along with the all-particle energy spectra,
are shown in Fig.~12 (lines and shaded areas) for the SIBYLL (left panel)
and QGSJET (right panel) interaction models.
The symbols in Fig.~12 show the all-particle spectra obtained by
KASCADE \cite{KASCADE2} from a 2-dimensional ($N_e,N_{\mu}$) unfolding
using an iterative method, and from GAMMA \cite{ICRC29b} using
an event-by-event method.
Also shown as error bars in the left panel of
Fig.~12 are extrapolations of the 
balloon and satellite data to the energy $E \simeq 10^6$ GeV,
computed using power-law approximations
to the available direct measurement data \cite{Wiebel}; these
remain in reasonable agreement with more recent balloon experiment
data \cite{JACEE,RUNJOB1}.  In this extrapolation, the $O$-like
group was assumed to include the elements $Z=3$--16, and the
$Fe$-like group the elements $Z=17$--28.\\
The expected EAS spectra and $N_{ch}(s)$ and $N_{ch}(N_{\mu})$ 
dependencies according to the solutions presented above are shown in 
Figs.~6--11 for the QGSJET (dashed lines) and SIBYLL (solid lines and 
shaded areas) interaction models. The vertical widths of the shaded areas
correspond to the error bars of the expected spectra, which are
comparable for the two interaction models.\\
It should be noted that the results obtained
in the framework of the QGSJET interaction model 
strongly depend on the number of examined functions,
which is not the case with the SIBYLL model.

\subsection{2-Dimensional approach} 
Using (1), parametrization (3) and the 2-dimensional EAS spectra 
\begin{displaymath}
F(\mathbf{q})\equiv
\frac{\d^2 F}{\d N_{ch}\d N_{\mu}} 
\end{displaymath}
we evaluated the parameters of the primary energy spectra by minimization
of the corresponding $\chi^2$ function (4), with $U=1$. The computations
were carried out with bin dimensions $\Delta\ln{N_{ch}}=0.15$ and
$\Delta\ln{N_{\mu}}=0.25$, for $\theta<30^\circ$ and $N_{ch}>5\cdot10^5$.
The resulting number of degrees of freedom ($n_{d.f.}$) for the $\chi^2$
minimization was equal to about 240.\\
The best-fit spectral parameters and 
corresponding values of $\chi^2/n_{d.f.}$ for both interaction
models are presented in Table~2.
\begin{table}   
\caption{\label{tab:table2}
Parameters of the primary energy spectra (3) 
from 2-dimensional approximations to the EAS data.
The scale factors $\Phi_A$ and particle rigidity $E_R$
respectively have units of (m$^2\cdot$ s $\cdot$ sr $\cdot$ TeV)$^{-1}$
and TV.}
\begin{center}
\begin{tabular}{lcc}
\hline
Parameters & SIBYLL&QGSJET\\
\hline
$\Phi_  H$ & $0.109\pm0.006$ & $0.198\pm0.006$\\
$\Phi_{He}$& $0.095\pm0.006$ & $0.028\pm0.005$\\
$\Phi_O   $& $0.050\pm0.006$ & $0.031\pm0.002$\\
$\Phi_{Fe}$& $0.017\pm0.002$ & $0.006\pm0.002$\\
$E_R      $& $2500\pm200$    & $4200\pm300$   \\
$\gamma_1$ & $2.70\pm0.005$ & $2.71\pm0.030$\\
$\gamma_2$ & $3.23\pm0.08$   & $3.23\pm0.09$\\
$\chi^2/n_{d.f.}$   & $1.2$ &$1.3$\\
\hline
\end{tabular}
\end{center}
\end{table}
The contributions to the total $\chi^2$ from each 2-dimensional 
bin $\mathbf{q}_i = (N_{ch},N_{\mu})$ at the minimum of the
$\chi^2$ function are shown in Figs.~13 and 14,
for the SIBYLL and QGSJET models respectively.
\begin{figure}   
\begin{center}
\includegraphics[width=8cm,height=8cm]{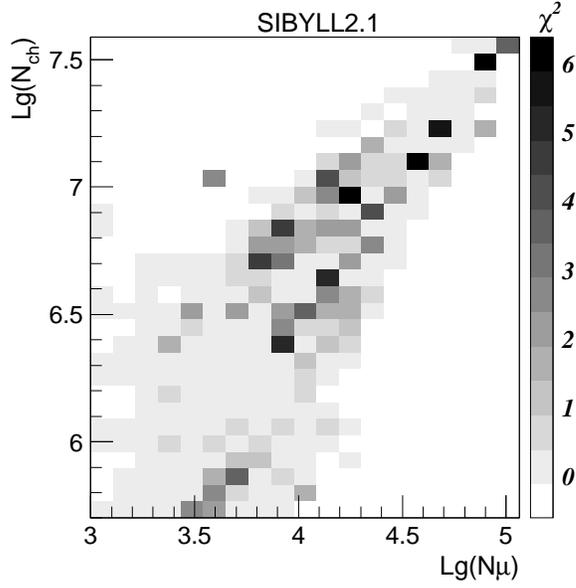}
\end{center}
\caption{Contributions to the total $\chi^2$ from each
($N_{ch},N_{\mu}$) bin, for the SIBYLL interaction model.}
\vspace{10mm}
\end{figure}
\begin{figure}   
\begin{center}
\includegraphics[width=8cm,height=8cm]{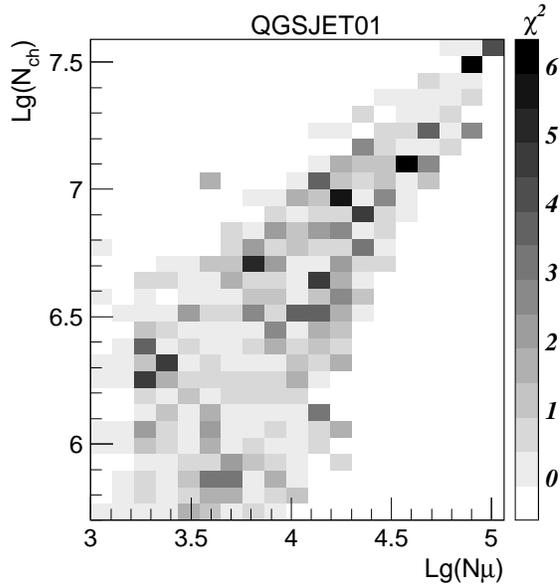}
\end{center}
\caption{Same as Fig.~13 for the QGSJET interaction model.}
\vspace{10mm}
\end{figure}

\subsection{4-Dimensional approach} 
The amount of 
information about the primary energy spectra contained
in the 4-dimensional spectrum of measured parameters
\begin{displaymath}
F(\mathbf{q})\equiv
\frac{\d^4 F}{\d N_{ch}\d N_{\mu}\d s\d\cos\theta}
\end{displaymath}
is obviously always greater than the information contained 
in the 2-dimensional ($N_{ch},N_{\mu}$) spectrum (Section 3.4)
or the cumulative amount of information contained 
in the combined spectra (Section 3.3).
The main difference with the latter case is due to
the inter-correlations between EAS parameters, which
can only be fully taken into account in such a
4-dimensional approach.\\
\begin{table}   
\caption{\label{tab:table3}
Parameters of the primary energy spectra (3) 
from the 4-dimensional analysis of the EAS data.
The scale factors $\Phi_A$ and particle rigidity $E_R$
respectively have units of (m$^2\cdot$ s $\cdot$ sr $\cdot$ TeV)$^{-1}$
and TV.}
\begin{center}
\begin{tabular}{lcc}
\hline
Parameters & SIBYLL&QGSJET\\
\hline
$\Phi_  H $ & $0.110\pm0.004$ & $0.190\pm0.002 $\\
$\Phi_{He}$ & $0.091\pm0.004$ & $0.023\pm0.003$\\
$\Phi_O   $ & $0.045\pm0.004$ & $0.038\pm0.002$\\
$\Phi_{Fe}$ & $0.030\pm0.002$ & $0.010\pm0.002$\\
$E_R      $ & $2300\pm230   $ & $3100\pm200   $\\
$\gamma_1 $ & $2.67\pm0.005  $ & $2.68\pm0.005$  \\
$\gamma_2 $ & $3.13\pm0.06  $ & $3.10\pm0.06$  \\
$\chi^2/n_{d.f.}$ & $2.1        $ & $2.1$          \\
\hline
\end{tabular}
\end{center}
\end{table}
On the basis of this 4-dimensional representation of the 
EAS data set, the simulated EAS database, and parameterization (3), 
equation (1) was solved by $\chi^2$ minimization, with $U=1$.
The computations were carried out with the following bin dimensions: 
$\Delta\ln{N_{ch}}=0.15$, 
$\Delta\ln{N_{\mu}}=0.25$,
$\Delta\sec{\theta}=0.05$,
and $\Delta s=0.15$ on the left and right hand side of $s^*=0.85$ 
and $\Delta s=0.3$ elsewhere. 
The number of degrees of freedom in this 4-dimensional 
approximation was equal to $1640$.
The values of spectral parameters (3) resulting from the solution
of the parameterized equation (1) are presented in Table~3
for the QGSJET and SIBYLL interaction models. 

\section{Discussion} 
As can be seen from Fig.~12 and Tables~1--3, the derived
primary energy spectra depend significantly on the interaction model,
and slightly on the approach (Sections 3.2--3.5) applied to solve the EAS 
inverse problem.
The derived abundances of primary nuclei at an energy $E\sim10^3$ TeV
in the framework of the SIBYLL model agree (in the range
of 1-2 standard errors) with the corresponding 
extrapolations of the balloon and satellite data \cite{Wiebel}, 
whereas the results derived with the QGSJET model point
toward a dominantly proton primary composition in the $10^3-10^5$
TeV energy range.\\
Although the derived formal accuracies of the spectral
parameters in Tables~1--3 are high, the corresponding $\chi^2$
values are large, which demands further discussion.
These large $\chi^2$ values do not necessarily imply
disagreement between the EAS data and the derived
primary energy spectra, but could be due to a number
of other possible causes. 
We believe that the most likely causes of the large $\chi^2$
values of our spectral fits are systematic uncertainties related
to the EAS simulations, in the interaction model or in the
computation of the detector response (Section~2.3),
and to the representation of the full cosmic ray composition
by a small number of simulated nuclear species.
We note that the inclusion of additional errors of
about $5-7\%$ in the $\chi^2$ functions (4)
will decrease the $\chi^2/n_{d.f.}\simeq2$ in Tables~1--3 to 
$\chi^2/n_{d.f.}\simeq1$.\\
We discuss in turn below a number of other possible causes and
related issues, especially the possibility that our spectral
parametrization is incorrect, in terms of the rigidity-dependent
knee energy or common spectral index.  We also consider briefly
the uncertainties in the reconstructed spectral parameters,
discuss possible issues with the convergence of the unfolding
method and the number of elemental groups, and present some
consistency checks on the simulated and experimental databases.

\subsection{Rigidity-dependent knee hypothesis}
A test of the spectral parametrization (3) was performed by 
evaluating the knee energies 
$E_k(A)$ independently for each primary nucleus
$A\equiv H,He,O,Fe$ simultaneously with the spectral
parameters $\Phi_A$,$\gamma_1$ and $\gamma_2$, using the
combined approximation method described above (Section 3.3).
The derived scale factors $\Phi_A$ and spectral indices 
$\gamma_1$ and $\gamma_2$ agreed with the data from Table~1
within errors, but they had somewhat larger uncertainties 
(typically by factors $\simeq 1.2 - 1.7$).
The derived knee energies versus nuclear charge ($Z$) are shown
in Fig.~15 for the SIBYLL
and QGSJET interaction models (symbols), along with the corresponding 
expected values (lines) according to the rigidity-dependent 
knee hypothesis from Table~1.
\begin{figure}   
\begin{center}
\includegraphics[width=8cm,height=8cm]{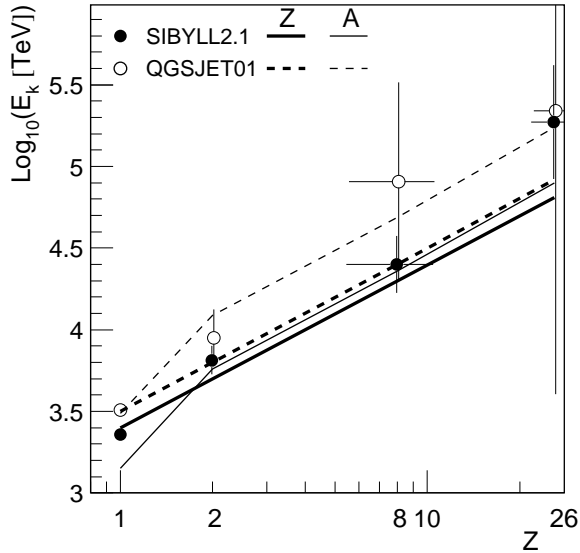}
\end{center}
\caption{Knee energies for each group of nuclei (symbols)
versus nuclear charge $Z$.
The expected values in the rigidity-dependent (thick lines)
and mass-dependent (thin lines) knee hypotheses are also shown,
 for the QGSJET
(dashed lines) and SIBYLL (solid lines) interaction models.}
\vspace{10mm}
\end{figure}
It can be seen from Fig.~15 that the independently adjusted
knee energies agree with the rigidity-dependent knee hypothesis.\\
We also examined the alternative mass-dependent knee hypothesis;
also shown as lines in Fig.~15 are the results of spectral fits
using combined approximations (Section 3.3), in which the hypothesis 
$E_k(A)=E_k\cdot A$, with $A$ the nuclear mass, was assumed.
The values of the $\chi^2_{\min}$ were practically the same as
in Table~1, but the derived value of the spectral parameter $\gamma_1$
tended to the range $2.59\pm0.02$, which is somewhat hard relative
to expectations from the balloon and satellite data
\cite{Wiebel,JACEE,RUNJOB1}.  Within the uncertainties of
our present analysis, our data are not in contradiction with this
$A$-dependent knee hypothesis; however, it clearly does not yield a
better agreement than our assumed rigidity-dependent hypothesis.

\subsection{Common spectral index hypothesis}
We attempted to similarly examine the possibility of independent
spectral indices $\gamma_{1,A}$ for each primary nucleus,
$A\equiv H,He,O,Fe$, but in that case encounter a difficulty.
The solution found by $\chi^2$ minimization when these parameters
are independent strongly depends on the initial values given
to the minimization algorithm, making a thorough exploration
of the multi-dimensional parameter space impractical, and the
results inconclusive.\\
Figure 16 shows the $\chi^2(\gamma_1)$ dependences for different
spectral hypotheses.  The thick solid line represents
the $\chi^2(\gamma_1)$ for parameterization (3), where the
spectral index is common to all primary species, obtained in
the combined approximation approach (Section 3.3) with the
SIBYLL interaction model.
\begin{figure}   
\begin{center}
\includegraphics[width=8cm,height=8cm]{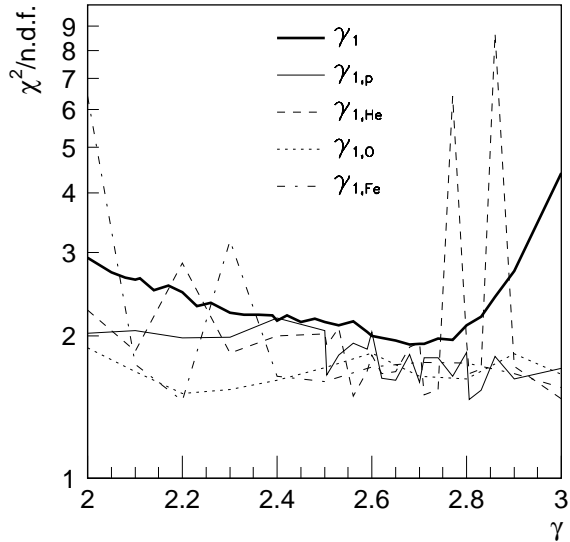}
\end{center}
\caption{  
$\chi^2_{\min}$ dependences versus the values of the fixed
parameter ($\gamma_1$) with randomly chosen initial values for
the adjustable parameters (see text for details).}
\vspace{10mm}
\end{figure}
The other lines show the corresponding dependences
$\chi^2(\gamma_{1,A})$ for individual nuclei, in the case
where the lower spectral indices $\gamma_{1,A}$ are independent
for each species.  In all cases, the value of the parameter
shown is held fixed, but values for all the other parameters
are obtained by minimization of (4), with initial values for
the minimization algorithm assigned randomly in a range of 
$\sim 10 - 20$\% around representative values for the
spectral parameters $\gamma_{1,A}$, $\gamma_2$ and 
$\log E_R$, and in a range of $\sim 50 - 100$\% for the
scale factors $\Phi_{A}$.
It is readily seen that while the curve $\chi^2(\gamma_1)$
for a common $\gamma_1$ shows a quite robust behavior, the
minima found for independent spectral indices strongly depend
on the initial values.\\
The shape of the $\chi^2(\gamma_1)$ curve for the parametrization
with equal spectral indices (3) may be used as an illustration
to examine the reliability of the uncertainties quoted in
Tables 1--3.  The minimization in all cases was performed using
the FUMILI algorithm \cite{FUMILI}, and the errors quoted were
obtained from the formal covariance matrix of the fit at the
$\chi^2$ minimum.  A more accurate estimation of the confidence
interval can be obtained from the intersection of the appropriate
level $\Delta \chi^2$ above the minimum $\chi^2$ value with a
curve such as the thick solid line in Fig.~16.  After normalizing
the errors such that $\chi^2 / n_{d.f.} \sim 1$, we find that the
actual confidence interval is somewhat wider than that obtained
from the formal uncertainty.  In general, our investigations
suggest that the derived formal errors tend to underestimate the
actual uncertainties in the spectral parameters by up to a factor
of two.

\subsection{Problem of uniqueness}
The example of independent spectral indices $\gamma_{1,A}$
illustrates a more general potential difficulty.
The EAS inverse problem is an 
ill-conditioned problem by definition, and unfolding of the 
corresponding integral
equations (1) does not ensure the uniqueness of the solutions.
Furthermore the EAS inverse problem implies the evaluation of
two or more unknown primary energy spectra from an
integral equation set of the Fredholm kind,
and this peculiarity has not been studied in detail.\\
Evidently, the solution cannot be considered unique
if a small change in the initial values of the iterative algorithm
for the minimization of (4) results in a significant change
(well beyond the formal uncertainties) of the solution spectra.
Using this test of uniqueness we
concluded that only the equality of the spectral indices for
all primary nuclei below the knee and the same equality of 
the spectral indices above the knee (parameterization (3)) result
in the unique solutions presented in Fig.~12 and Tables~1--3.

\subsection{Number of elemental groups}
The evaluations of primary spectra for pure $H$, pure $He$ and mixed
($H,He$), ($H,He,O$) and ($H,He,Fe$) compositions in parameterization
(3) also were examined using the 2-dimensional approach (Section 3.4).
The corresponding $\chi^2/n_{d.f.}$ values were respectively equal to
$44.5, 35.3, 10.0, 1.8$ and $2.5$ for the SIBYLL interaction model
and $11.5, 141, 4.0, 2.7$ and $2.0$ for the QGSJET model.
The results for mixed $H,He,O$ and $Fe$ primary composition
are presented in Table~2.  It is readily seen that the data
cannot be adequately represented with less than the four considered
types of primary nuclei.\\
Examining these results we can conclude that increasing the
number of considered primary nuclei in our parameterized
inverse approach increases the accuracy of the solutions.
This effect indirectly supports the validity of our 
parametrization with equal spectral indices. 
If our assumption of the equality of the spectral indices
was invalid, we would not expect the $\chi^2$ to improve
so effectively with increasing number of nuclear species.

\subsection{Consistency of the solutions}
The agreement of the data presented in Tables~1 and 3 with our
preliminary results \cite{GAMMA05,ICRC29a}, which were obtained
with significantly fewer (half as many) simulated showers, suggests
that the size of the simulated database is not a problem.\\
\begin{figure}   
\begin{center}
\includegraphics[width=8cm,height=8cm]{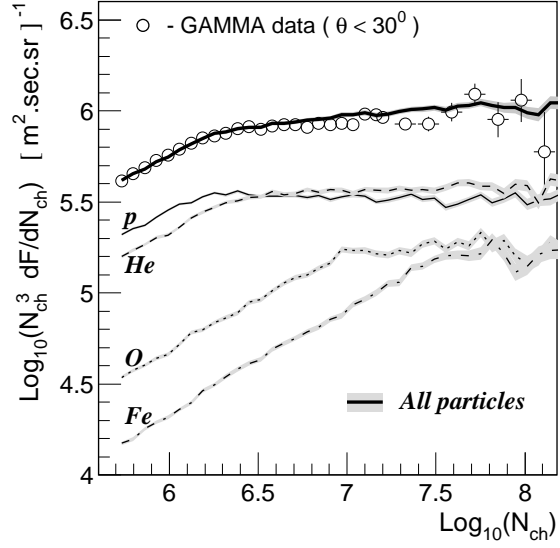}
\end{center}
\caption{EAS size spectrum with an enlarged shower core selection 
criterion ($R<50$~m) (symbols),
and expected shower spectra for each of the primary nuclei  
and the mixed composition (lines and shaded areas), with
parameters from Table~1 and
for the SIBYLL interaction model.}
\vspace{10mm}
\end{figure}
\begin{figure}   
\begin{center}
\includegraphics[width=8cm,height=8cm]{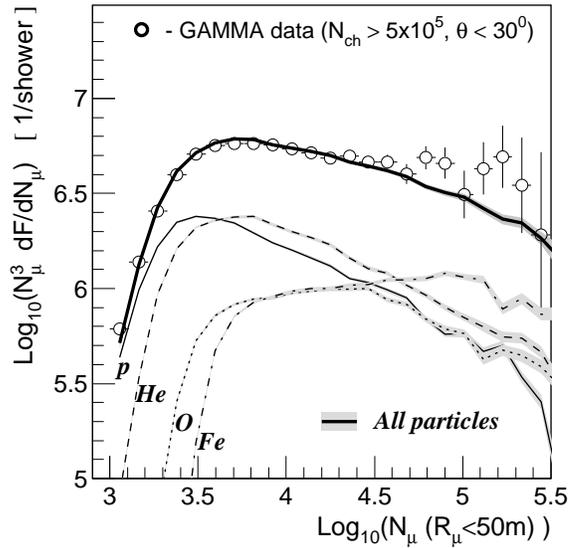}
\end{center}
\caption{Same as Fig.~16 for the EAS truncated muon size spectra.}
\vspace{10mm}
\end{figure}
A further check of the consistency of the GAMMA facility
EAS data with the derived solutions is shown in Figs.~17 and 18,
which display the EAS size and truncated muon size spectra
(symbols) for an enlarged core selection criterion of $R<50$ m.
This is twice as large as the selection radius of the EAS
data in Figs.~6--7, and resulted in about four times the number
of selected showers.  
The lines and shaded areas in Figs.~17--18 correspond to the expected 
(forward folded) EAS spectra with the parameters of primary energy
spectra (3)  
from Table~1 for the SIBYLL interaction model; the corresponding
expected shower spectra for each of primary nuclei are also shown.

\section{Conclusion} 
The consistency of the results obtained by the GAMMA experiment
(Figs.~6--11,16--17), at least up to $N_{ch}\simeq10^7$, with the
corresponding predictions in the
framework of the hypothesis of a rigidity-dependent knee in the primary
energy spectra and the validity of the SIBYLL or QGSJET 
interaction models points toward the following conclusions:
\begin{itemize}
\item A rigidity-dependent steepening of primary energy spectra 
in the knee region
(expression 3) describes the EAS data of the GAMMA experiment at 
least up 
to $N_{ch}\simeq10^7$ with an average accuracy $<10\%$, with particle 
magnetic rigidities
$E_R\simeq2500\pm200$ TV (SIBYLL) and $E_R\simeq3100-4200$ TV
(QGSJET).
The corresponding spectral power-law indices are
$\gamma_1=2.68\pm0.02$ and $\gamma_2=3.10-3.23$
below and above the knee respectively, and the element group
scale factors $\Phi_A$ are given in Tables~1--3.
\item The abundances and energy spectra obtained for primary $p$, 
$He$, $O$-like and $Fe$-like nuclei depend on the interaction 
model. 
The SIBYLL interaction model is preferable in terms of consistency
of the extrapolations of the derived primary spectra (Fig.~12)
with direct measurements in the energy range of 
satellite and balloon experiments \cite{Wiebel,JACEE,RUNJOB1}.
\item The derived all-particle energy spectra depend only
weakly on the interaction model.  They are compatible with
independent measurements of this spectrum.
\item An anomalous behavior of the EAS muon size and density
spectra (Fig.~5b, Figs.~11 and 18)
and the EAS age parameter (Fig.~10) for EAS size $N_{ch}>10^7$ is observed.
A similar behavior of the EAS age parameter has previously been observed in 
\cite{MAKET,Norikura}.
The observed behavior of the muon size and density 
spectra may be
related to the excess of high-multiplicity cosmic muon events
detected by the ALEPH and DELPHI experiments \cite{Taylor,LeCoultre}.
\end{itemize}

\section*{Acknowledgments} 
We wish to thank Anatoly Erlykin for helpful discussions,
and an anonymous referee for suggestions which considerably improved
the paper.
We are grateful to all of our colleagues at the Moscow Lebedev 
Institute and the Yerevan Physics Institute who took part 
in the development and exploitation of the GAMMA array.\\
This work has been partly supported by research grant No 090 
from the Armenian government,
by CRDF grant AR-P2-2580-YE-04, and by the ``Hayastan''
All-Armenian Fund and the ECO-NET project 12540UF in France.

\end{document}